\documentclass{article}
    \usepackage{amssymb}
    \usepackage{amsmath,amsfonts,amsthm}
    \usepackage{fullpage} 

\usepackage{definitions}
\usepackage{ifthen}
\usepackage{xspace}
\usepackage{algorithm}
\usepackage{algorithmic}

\renewcommand{\Equation}[2]{
\begin{align*} #1 & =  #2 \end{align*}
}

\renewcommand{\LEquation}[2]{
\begin{align*} #1 & \leq  #2 \end{align*}
}

\newcommand{\remove}[1]{}

\renewcommand{\Abs}[1]{\ensuremath{\left| #1 \right|}\xspace}
\renewcommand{\Expect}[1]{\ensuremath{{\rm E}\!\left[#1\right]}\xspace}

\newcommand{\eps}{\epsilon}

\newcommand{\dx}{{\rm d}x}
\newcommand{\dy}{{\rm d}y}
\newcommand{\dt}{{\rm d}t}

\newcommand{\unto}{{\rm unto}}
\newcommand{\unex}{{\rm unex}}

\newcommand{\vis}{{\rm vis}}
\newcommand{\obs}{{\rm obs}}
\newcommand{\hg}{{}_{2}F_{1}}
\newcommand{\Ei}{{\rm Ei}}
\newcommand{\e}{{\rm e}}
\newcommand{\p}{\partial}
\newcommand{\dmin}{{d_{\rm min}}}

\newcommand{\whp}{w.h.p.\xspace}
\newcommand{\wohp}{w.o.h.p.\xspace}
\newcommand{\bin}{{\rm Bin}}
\newcommand{\ie}{i.e.,\xspace}
\newcommand{\reasonable}{reasonable\xspace}
\newcommand{\Psib}{P_{\rm sib}}

\newcommand{\cunto}[1][]{\ensuremath{%
\ifthenelse{\equal{#1}{}}{c_\unto}{c_\unto(#1)}}\xspace}
\newcommand{\Cunto}[1][]{\ensuremath{%
\ifthenelse{\equal{#1}{}}{C_\unto}{C_\unto(#1)}}\xspace}
\newcommand{\pvis}[1][]{\ensuremath{%
\ifthenelse{\equal{#1}{}}{p_\vis}{p_\vis(#1)}}\xspace}
\newcommand{\Pvis}[1][]{\ensuremath{%
\ifthenelse{\equal{#1}{}}{P_\vis}{P_\vis(#1)}}\xspace}
\newcommand{\Qvis}[2]{\ensuremath{{\rho_{{#1,#2}}}}\xspace}
\newcommand{\vunto}[2][]{\ensuremath{%
\ifthenelse{\equal{#1}{}}{v_{\unto,#2}}{a_{#2}#1^{#2}n}}\xspace}
\newcommand{\Vunto}[2][]{\ensuremath{%
\ifthenelse{\equal{#1}{}}{V_{\unto,#2}}{V_{\unto,#2}(#1)}}\xspace}
\newcommand{\PUNTO}[1][]{\ensuremath{%
\ifthenelse{\equal{#1}{}}{P_\unto}{P_\unto(#1)}}\xspace}
\newcommand{\Punto}[2][]{\ensuremath{%
\ifthenelse{\equal{#1}{}}{P_{\unto,#2}}{P_{\unto,#2}(#1)}}\xspace}
\newcommand{\cunex}[1][]{\ensuremath{%
\ifthenelse{\equal{#1}{}}{c_\unex}{c_\unex(#1)}}\xspace}
\newcommand{\Cunex}[1][]{\ensuremath{%
\ifthenelse{\equal{#1}{}}{C_\unex}{C_\unex(#1)}}\xspace}
\newcommand{\Queue}[1][]{\ensuremath{%
\ifthenelse{\equal{#1}{}}{Q}{Q(#1)}}\xspace}
\newcommand{\queue}[1][]{\ensuremath{%
\ifthenelse{\equal{#1}{}}{q}{q(#1)}}\xspace}
\newcommand{\QueueSet}[1][]{\ensuremath{%
\ifthenelse{\equal{#1}{}}{{\cal Q}}{{\cal Q}(#1)}}\xspace}
\newcommand{\VisDeg}[2][]{\ensuremath{%
\ifthenelse{\equal{#1}{}}{A^{\obs}_{#2}}{B_{#2}(#1)}}\xspace}
\newcommand{\bvec}{\ensuremath{\mathbf{b}}\xspace}

\def\more-auths{%
\end{tabular}
\begin{tabular}{c}}

\begin{document}

\title{On the Bias of Traceroute Sampling: or, Power-law Degree Distributions in Regular Graphs}

\author{
\hspace{11mm} Dimitris Achlioptas \hspace{11mm} \\
	Microsoft Research \\
	Microsoft Corporation \\
	Redmond, WA 98052 \\
	\tt{optas@microsoft.com}	
\and
Aaron Clauset\\ 
Department of Computer Science \\
University of New Mexico \\
Albuquerque, NM 87131 \\
\tt{aaron@cs.unm.edu}
\and
David Kempe
\\
Department of Computer Science \\
University of Southern California \\
Los Angeles, CA 90089 \\
\tt{dkempe@usc.edu}
\and
Cristopher Moore
\\
Department of Computer Science \\
University of New Mexico \\
Albuquerque, NM 87131 \\
\tt{moore@cs.unm.edu}
}

\maketitle

\begin{abstract}
Understanding the structure of the Internet graph is a crucial step
for building accurate network models and designing efficient 
algorithms for Internet applications.
Yet, obtaining its graph structure is a surprisingly
difficult task, as edges cannot be explicitly queried. Instead,
empirical studies rely on \todef{traceroutes} to build what are
essentially single-source, all-destinations, shortest-path trees. 
These trees only sample a fraction of the network's edges, and
a recent paper by Lakhina et al.~found empirically that the resuting sample 
is intrinsically biased. For instance, 
the observed degree distribution under traceroute sampling 
exhibits a power law even when the underlying degree distribution
is Poisson.

In this paper, we study the bias of traceroute sampling systematically, and, for 
a very general class of underlying degree distributions, calculate the likely 
observed distributions explicitly. To do this, we use a continuous-time 
realization of the process of exposing the BFS tree of a random graph
with a given degree distribution, 
calculate the expected degree distribution of the tree, and 
show that it is sharply concentrated.  As example applications
of our machinery, we show how traceroute sampling finds power-law 
degree distributions in both $\delta$-regular and Poisson-distributed random graphs.
Thus, our work puts the observations of Lakhina et al. on a rigorous footing, 
and extends them to nearly arbitrary degree distributions.
\end{abstract}

\remove{
\vspace{1mm}
\noindent
{\bf Categories and Subject Descriptors:} F.2.2, G.2.1

\vspace{1mm}
\noindent
{\bf General Terms:} Theory

\vspace{1mm}
\noindent
{\bf Keywords:} Internet mapping, power laws, random graphs
}

\section{Introduction}
Owing to the great importance of the Internet as a medium for
communication, a large body of recent work has focused on its
topological properties. Perhaps most famously, Faloutsos et
al.~\cite{faloutsos:faloutsos:faloutsos} exhibited a power-law degree
distribution in the Internet graph at the router level (\ie the level at 
which the Internet Protocol (IP) operates).
Similar results were obtained in
\cite{govindan:tangmunarunkit,barford:bestavros:byers:crovella}. Based
on these and other topological studies, it is widely believed that
the Internet's degree distribution has a power-law form with exponent 
$2 < \alpha < 3$, \ie the fraction $a_k$ of vertices with degree $k$ is
proportional to $k^{-\alpha}$. These results have motivated both the
search for natural graph growth models that give similar
degree distributions (see for instance
\cite{fabrikant:koutsoupias:papadimitriou:HOT}) and research into the
question of how the topology might affect the performance of Internet algorithms and
mechanisms (for instance
\cite{mihail:papadimitriou:saberi:connectivity}).

However, unlike graphs such as the World Wide Web
 \cite{kleinberg:kumar:raghavan:rajagopalan:tomkins} in which links from 
each site can be readily observed, the physical connections between routers on the 
Internet cannot be queried directly. Without explicitly knowing which routers 
are connected, how can one obtain an accurate map of the Internet? Internet 
mapping studies typically address this issue by sampling the network's 
topology using \todef{traceroutes}: packets are sent across the network 
in such a way that their paths are annotated with the IP addresses of the routers that forward them. 
The union of many such paths then forms a partial map of the Internet. 
While actual routing decisions involve multiple protocols 
and network layers, it is a common assumption that the packets
follow shortest paths between their source and destination, and recent studies 
show that this is not far from the truth~\cite{latapy:internet-routes}.

Most studies, including the one on which \cite{faloutsos:faloutsos:faloutsos} 
is based, infer the Internet's topology from the union of traceroutes 
from a single \todef{root} computer to a larger number of (or all)
other computers in the network.  If each edge has unit cost plus a small random term, 
the union of these shortest paths is a BFS
tree.\footnote{Several studies, including 
  \cite{barford:bestavros:byers:crovella,pansiot:grad:routers}, have
  used traceroutes from multiple sources. However, the number of
  sources used is quite small (to our knowledge, at most 12).}
This model of the sampling process is admittedly an idealization for several 
reasons.  First of all, most empirical studies 
only use a subset of the valid IP addresses as destinations.  Secondly, 
for technical reasons, some routers may not respond to traceroute queries. 
Thirdly, a single router may annotate different traceroutes with different IP addresses, a problem known as {\em aliasing}.  These issues are known to introduce noise into the 
measured topology~\cite{amini:shaikh:issues, chen:chang:origins}.

However, as Lakhina et al.~\cite{lakhina:byers:crovella:xie} recently pointed out, 
traceroute sampling has a more fundamental bias, 
one which is well-captured by the BFS idealization.  Specifically, 
in using such a sample to represent the network, one tacitly assumes that
the sampling process is unbiased with respect to the parameters under consideration, such as node
degrees.  However, an edge is much more likely to be {\em visible}, \ie included 
in the BFS tree, if it is close to the root.  Moreover, since in a random graph,  
high-degree vertices are more likely to be encountered early on in the BFS tree, 
they are sampled more accurately than low-degree vertices.  
Indeed, \cite{lakhina:byers:crovella:xie} 
showed empirically that for Erd\H{o}s-R\'enyi random graphs
$G(n,p)$~\cite{erdos:renyi:gnp}, which have a Poisson degree
distribution, the observed degree distribution under traceroute
sampling follows a power law,  
and this has been verified analytically by 
Clauset and Moore~\cite{clauset:moore:internet-mapping}.
In other words, the bias introduced by traceroute sampling can make
power laws appear where none existed in the underlying graph!
Even when the underlying graph actually does have a power-law degree
distribution $k^{-\alpha}$, Petermann and De Los 
Rios~\cite{petermann:delosrios:exploration} 
and Clauset and Moore~\cite{clauset:moore:internet-mapping} 
showed numerically that traceroute sampling can significantly underestimate its 
exponent $\alpha$.

This inherent bias in traceroute sampling (along with the fact that no
alternatives are technologically feasible at this point) raises the following
interesting question: Given the true degree distribution \SET{a_k} of
the underlying graph, can we predict the degree distribution that will
be observed after traceroute sampling? Or, in pure graph-theoretic
terms: can we characterize the degree distribution of a BFS tree for a
random graph with a given degree distribution?

Our answers to these questions quantify precisely the bias
introduced by traceroute sampling, while verifying formally the empirical
observations of Lakhina et al.~\cite{lakhina:byers:crovella:xie}.
In addition, they can be considered as a significant first step toward 
a much more ambitious and ultimately more practical goal of inferring
the true underlying distribution of the Internet from the biased observation.

\subsubsection*{Our Results}

Our main result in this paper is Theorem \ref{thm:genfunc}, which
explicitly characterizes the observed degree distribution as a
function of the true underlying distribution, to within sharp
concentration. When we say that \SET{a_k} is a degree distribution,
what we mean precisely is that the graph contains $a_k \cdot n$ nodes of degree
$k$. In proving the result, we restrict our attention to underlying
distributions which are ``not too heavy-tailed,'' and in which all
nodes have degree at least 3:

\begin{definition} \label{def:reasonable}
A degree distribution \SET{a_k} is \todef{\reasonable} if $a_k = 0$ for
$k < 3$, and there exist constants $\alpha > 2$ and $C > 0$ such that
$a_k < C \cdot k^{-\alpha}$ for all $k$.
\end{definition}

The requirement that the degree distribution be bounded by a power law
$k^{-\alpha}$ with $\alpha > 2$ is made mostly for technical
convenience.
Among other things, it implies that 
the mean degree $\delta = \sum_k k a_k$ of the graph is finite 
(although the variance is infinite for $\alpha \le 3$). 
Note that this requirement is consistent with the conjectured range 
$2 \le \alpha \le 3$ for the Internet~\cite{faloutsos:faloutsos:faloutsos,govindan:tangmunarunkit}.
The requirement that the minimum degree be at least $3$ implies,
through a simple counting argument, that the graph is 
\whp\ connected.\footnote{We say that a sequence of events
  \Event[n]{E} occurs \todef{with high probability (\whp)} if
  $\Prob{\Event[n]{E}} = 1-o(1)$ as $n \to \infty$, and \todef{with
    overwhelmingly high probability (\wohp)} if 
   $\Prob{\Event[n]{E}} = 1-o(n^{-c})$ for all $c$.  Note that by the
   union bound, the conjunction of a polynomial number of events, each
   of which occurs \wohp, occurs \wohp}
This is convenient since it ensures that the breadth-first tree
reaches the entire graph.  However, as we discuss below, this
requirement can be relaxed, and in the case of disconnected graphs
such as $G(n,p=\delta/n)$, we can indeed analyze the breadth-first tree
built on the giant component.

In order to speak precisely about a random (multi)graph with a given
degree sequence, we will use the \todef{configuration 
  model}~\cite{bollobas:random-graphs}: for each
vertex of degree $k$, we create $k$ \todef{copies}, and then define
the edges of the graph according to a uniformly random matching on
these copies. Our main result can then be stated as follows:

\begin{theorem} \label{thm:genfunc}
Let \SET{a_j} be a \reasonable degree sequence.
Let $G$ be a random multigraph with degree distribution \SET{a_j}, and
assume that $G$ is connected.  Let $T$ be a breadth-first tree on $G$,
and let $A^\obs_j$ be the number of vertices of degree $j$ in $T$. 
Then, there exists a constant $\zeta > 0$ such that 
with high probability,
$\Abs{A^\obs_j - a^\obs_j n} < n^{1-\zeta}$ for all $j$, where
\begin{align*}
a^\obs_{m+1}
& = \sum_i a_i \Bigg[ \int_0^1 i t^{i-1} \,{i-1 \choose m} \,\pvis[t]^m
\,(1-\pvis[t])^{i-1-m} \, \dt \Bigg] \enspace,\\
\pvis[t]
& = \frac{1}{\sum_j j a_j t^j} \sum_k k a_k t^k \left( 
\frac{\sum_j j a_j t^j}{\delta t^2} \right)^{\!k} \enspace.
\end{align*}
We can use the notion of generating functions
\cite{wilf:generatingfunctionology} to obtain a more concise
expression of all $a^\obs_{m+1}$ as follows: 
if $g(z) = \sum_{j=0}^\infty a_j z^j$, then 
$a_j^\obs$ is the coefficient of $z^j$ in
\begin{align}
 g^\obs(z) & =  z \int_0^1 g'\!\left[ t-\frac{(1-z)}{g'(1)} \,
   g'\!\left( \frac{g'(t)}{g'(1)}\right)\right] \, \dt  \enspace.
\label{eq:gobs}
\end{align}
\end{theorem}

The bulk of this paper, namely Sections 2--5, is devoted to the proof of
Theorem~\ref{thm:genfunc}. In Section~\ref{sec:examples}, we apply our
general result to $\delta$-regular graphs and graphs with Poisson degree
distributions. In both cases, we find that the observed degree
distribution follows a power law $k^{-\alpha}$ with exponent $\alpha = 1$. In the
case of Poisson degree distributions, our work thus subsumes the work
of Clauset and Moore \cite{clauset:moore:internet-mapping}.

The proof of this result is based on a process which gradually
discovers the BFS tree (see Section~\ref{sec:process}). By mapping it
to a continuous-time process analogous to Kim's Poisson cloning
model~\cite{kim:poisson-cloning}, we can avoid explicitly tracking the
(rather complicated) state of the FIFO queue that arises in the
process, and in particular the complex relationship between a vertex's
degree and its position in the queue. This allows us to calculate the
expected degree distribution to within $o(1)$ in
Section~\ref{sec:expected}. In Section~\ref{sec:genfunc}, we see how
these calculations can be rephrased in terms of generating
functions, to yield the alternate formulation of Theorem
\ref{thm:genfunc}. The concentration part of the result, 
in Section~\ref{sec:concentration}, analyzes a different, and much
more coarse-grained, view of the process. By carefully conditioning on
the history of the process, we can apply a small number of
Martingale-style bounds to obtain overall concentration.

\section{A Continuous-Time Process}
\label{sec:process}


\subsection{Breadth-First Search}
We can think of the breadth-first tree as being built one vertex at a
time by an algorithm that explores the graph.  At each step, every
vertex in the graph is labeled \todef{explored}, \todef{untouched}, or
\todef{pending}.  A vertex is explored if both it and its neighbors are
in the tree; untouched if it is still outside the tree; and pending if
it is on the boundary of the tree, \ie it may still have untouched
neighbors.  Pending vertices are kept in a queue \QueueSet, so that
they are explored in first-in, first-out order.  The process is
initialized by labeling the root vertex pending, and all other
vertices untouched. Whenever a pending vertex is popped from \QueueSet
and explored, all of its currently untouched neighbors are appended to
\QueueSet, and the connecting edges are \todef{visible}. On the other
hand, edges to neighbors that are already in the queue are not
visible.

For the analysis in this paper, it is convenient to think of the
algorithm as exploring the graph one copy at a time, instead of one
node at a time. The queue will then contain copies instead of
vertices.  At each step, the partner $v$ of the copy $u$ at the head
of the queue is \todef{exposed}, and both of them are removed from the
matching. Also, all of $v$'s siblings are added to the queue, unless
they were in the queue already. 
(We refer to two copies of the same vertex as \todef{siblings}.) 
We will say that an unexposed copy is \todef{enqueued} if it is in
\QueueSet, and \todef{untouched} if it is not.  Thus,  
a copy is untouched if its vertex is, and 
enqueued if its vertex is pending and the edge incident to it has
yet to be explored.  Formally, the breadth-first search then looks as follows:
\begin{algorithm}
\caption{Breadth-First Search at the Copy Level} \label{alg:BFS}
\begin{algorithmic}[1]
\WHILE{\QueueSet is nonempty} 
\STATE Pop a copy $u$ from the head of \QueueSet 
\STATE Expose $u$'s partner $v$
\IF{$v$ is untouched}
\STATE Add the edge $(u,v)$ to $T$
\STATE Append $v$'s siblings to \QueueSet
\ELSE
\STATE Remove $v$ from \QueueSet
\ENDIF
\ENDWHILE
\end{algorithmic}
\end{algorithm}

An edge will be visible and included in $T$ if, at the time one of its endpoints
reaches the head of the queue, the other endpoint is still untouched.

\subsection{Exposure on the fly}
Because $G$ is a uniformly random multigraph conditioned on its
degree sequence, the matching on the copies is {\em uniformly random}.
By the principle of deferred decisions~\cite{motwani:raghavan}, we can
define this matching ``on the fly,'' choosing $u$'s partner $v$
uniformly at random from among all the unexposed copies at the time.

One way to make this random choice is as follows.  At the outset,
each copy is given a real-valued index $x$ chosen uniformly at
random from the unit interval $[0,1]$.  Then, at each step, $u$'s
partner $v$ is chosen as the unexposed copy with the the largest
index.
Thus, it is convenient to think of the algorithm as taking place in
continuous time, where $t$ decreases from $1$ to $0$: at time $t$,
the copy at the head of the queue is matched with the unexposed copy of
index $t$.  Since the indices of $v$'s siblings are uniformly random,
while conditioned on being less than $t$, this approach maintains the
following powerful kind of uniform randomness: at time $t$, the
indices of the unexposed copies, both inside and outside the queue,
are uniformly random in $[0,t)$.

We define the \todef{maximum index} of a vertex to be the maximum of
all its copies' indices. At any time $t$, the untouched vertices
are precisely those whose maximum index is less than $t$, and the
explored or pending vertices (whose copies are explored or enqueued)
are those whose maximum index is greater than $t$.  This observation
allows us to carry out an explicit analysis without having to track
the (rather complicated) state of the system as a function of time.

At a given time $t$, let \Cunex[t] and \Cunto[t] denote the
number of unexposed and untouched copies, and let \Vunto[t]{j} denote
the number of untouched vertices of degree $j$; 
note that $\Cunto[t] = \sum_j j \Vunto[t]{j}$.  
We start by calculating the expectation of
these quantities.  The probability that a vertex of degree $j$
has maximum index less than $t$ is exactly $t^j$; therefore,
$\Expect{\Vunto[t]{j}} = a_j t^j n$, 
and 
\begin{equation} \label{eq:unto} 
\begin{array}{lclcl}
\Expect{\Cunto[t]} & = & \sum_j j a_j t^j n & =: & \cunto[t] \cdot n \enspace . 
\end{array} \end{equation}

To calculate \Expect{\Cunex[t]}, recall that the copy at the head of the queue
has a uniformly random index conditioned on being less than $t$.  Therefore, 
the process forms a matching on the list of indices as follows: take the
indices in decreasing order from $1$ to $0$, and at time $t$ match the index
$t$ with a randomly chosen index less than $t$.  This creates a uniformly random 
matching on the $\delta n$ indices.  Now, note that a given index is still remaining 
at time $t$ if both it and its partner are less than $t$, and since 
the indices are uniformly random in $[0,1]$ the probability of this is $t^2$.  
Thus, the expected number of indices remaining at time $t$ is 
\begin{equation} \label{eq:unex}
\begin{array}{lclcl}
\Expect{\Cunex[t]} 
 & = & \delta t^2 n
 & =: & \cunex[t] \cdot n \enspace . 
\end{array} \end{equation}

The following lemma shows that \Vunto[t]{j}, \Cunto[t] and \Cunex[t]
are concentrated within $o(n)$ of their expectations throughout the
process. 
Note that we assume here that the graph $G$ is connected, since
otherwise, the process is not well-defined for all $t \in [0,1]$.
\begin{lemma} \label{lem:copies} 
Let \SET{a_j} be a \reasonable degree distribution, and assume that $G$
is connected. 
Then, for any constants $\beta < \min(\half, \frac{\alpha-2}{2})$ and
$\epsilon > 0$, the following hold simultaneously for all $t \in [0,1]$ and
for all $j < n$, \wohp:
\begin{align*}
\Abs{\Vunto[t]{j} - a_j t^j \cdot n} & <  n^{1/2+\epsilon}\\
\Abs{\Cunex[t] - \cunex[t] \cdot n} & <  n^{1/2+\epsilon}\\
\Abs{\Cunto[t] - \cunto[t] \cdot n} & <  n^{1-\beta},
\end{align*}
where \cunto[t] and \cunex[t] are given by~\eqref{eq:unto},~\eqref{eq:unex}.
\end{lemma}

Note that this concentration becomes weaker as $\alpha \to 2$, since then $\beta \to 0$.

\begin{proof}
Our proof is based on the following form of the Hoeffding
Bound~\cite{hoeffding,mcdiarmid:concentration-survey}:

\begin{theorem}[Theorem 3 from \cite{mcdiarmid:concentration-survey}]
\label{thm:hoeffding}
If $X_1, \ldots, X_k$ are independent, non-negative random variables
with $X_i \leq b_i$ for all $i$, and $X = \sum_i X_i$, then for any
$\Delta \geq 0$:
\LEquation{\Prob{\Abs{X - \Expect{X}} \geq \Delta}}{2\e^{-2\Delta^2/\sum_i b_i^2} \enspace .}
\end{theorem}

First, \Vunto[t]{j} is a binomial random variable distributed as
$\bin(a_j n,t^j)$.  By applying Theorem~\ref{thm:hoeffding} to $a_j n$
variables bounded by $1$, the probability that \Vunto[t]{j} differs by
$\Delta = n^{1/2+\epsilon}$ from its expectation is at most 
$2\e^{-2n^{2\epsilon} / a_j} \leq \e^{-n^{2\epsilon}}$. 
Thus, at each individual time $t$ and for each $j$, 
the stated bound on \Vunto[t]{j} holds \wohp 

We wish to show that this bound holds \wohp\ for all $j$ and all $t$,
\ie that the probability that it is violated for any $t$ and any $j$
is  $o(n^{-c})$ for all $c$. Notice that the space of all times $t$ is
infinite, so we cannot take a simple union bound. Instead, we divide
the interval $[0,1]$ into sufficiently small discrete subintervals,
and take a union bound of those.
Let $m = \sum_j j a_j n = \delta n$ be the total number of copies,
where $\delta$ is the mean degree (recall that $\delta$ is finite,
because \SET{a_j} is \reasonable).  We divide the unit interval
$[0,1]$ into $m^b$ intervals of size $m^{-b}$, where $b$ will be set below.
By a union bound over the ${m \choose 2}$ pairs of copies,
with probability at least $1-m^{2-b}$, each interval contains the index 
of at most one copy, and therefore at most one event of the queue
process. Conditioning on this event,
\Vunto[t]{j} changes by at most $1$ during each interval, so if
$\Vunto[t]{j}$ is close to its expectation at the boundaries of each interval,
it is close to its expectation for all $t \in [0,1]$. In
addition, we take a union bound over all $j$.  The probability that
the stated bound is violated for any $j$ in any 
interval is then at most
\Equation{n \left( m^b \,\e^{-n^{2\eps}} + m^{2-b} \right)}{O(n^{3-b}) \enspace ,}
which is $o(n^{-c})$ if $b > c+3$.

For the concentration of \Cunex[t], we notice that unexposed copies
come in matched pairs, both of which have index less than $t$.
Therefore, \Cunex[t] is twice a binomial random 
variable distributed as $\bin(\sum_j j a_j n/2, t^2)$.  
Applying Theorem~\ref{thm:hoeffding} with $\Delta = n^{1/2+\epsilon}$
gives the result for fixed $t$, and taking a union bound over
$t$ as in the previous paragraph shows the concentration of \Cunex[t].

To prove concentration of \Cunto[t] for fixed $t$, we 
let $X_i$ be the number of copies of node $i$ that are untouched at
time $t$. Then, $\Cunto[t] = \sum_i X_i$, and the denominator in the
exponent for the bound of Theorem~\ref{thm:hoeffding} is
\begin{align*}
\sum_i b_i^2 
\;=\;  \sum_j j^2 a_j n
\;<\;  Cn \sum_j j^{2-\alpha}
\;<\;  \left\{ \begin{array}{ll}
O(n^{4-\alpha}) & \alpha < 3 \\
O(n \log n) & \alpha = 3 \\
O(n) & \alpha > 3 
\end{array} \right.  
\end{align*}
Hence, by Theorem~\ref{thm:hoeffding}, whenever 
$\beta < \min(\half, \frac{\alpha-2}{2})$, we 
obtain that $\Abs{\Cunto[t]-\Expect{\Cunto[t]}} \leq n^{1-\beta}$ \wohp
A union bound over $t$ as before 
completes the proof.
\end{proof}

\section{Expected degree distribution}
\label{sec:expected}

In this section, we begin the proof of Theorem~\ref{thm:genfunc} by
analyzing the continuous-time process defined in Section~\ref{sec:process},
and calculating the expected degree distribution of the tree $T$.

By linearity of expectation, the expected number of vertices 
of degree $j$ in $T$ is the sum, over all vertices $v$, of the probability 
that $j$ of $v$'s edges are visible.  Consider a given vertex $v$ of degree $i$.   
It is touched when its copy with maximum 
index is matched to the head of the queue, at which time
its $i-1$ other copies join the tail of the queue.  If $m$ of these
give rise to visible edges, then $v$'s degree in $T$ will be $m+1$, 
namely these $m$ outgoing edges plus the edge connecting $v$ 
back toward the root of the tree.  

Let \Qvis{i}{m} denote the probability of this event, \ie 
that a vertex of degree $i$ has $m$ copies that give rise 
to visible edges.  Then the expected degree distribution is given by 
\begin{equation}
\Expect{\VisDeg{m+1}} = n \sum_i a_i \Qvis{i}{m} \enspace . 
\label{eq:visdeg}
\end{equation}
Moreover, let $\Qvis{i}{m}(t)$ denote the probability of this event 
given that $v$ has maximum index $t$.  Then, since $t$ is the 
maximum of $i$ independent uniform variables in $[0,1]$, its
probability distribution is ${\rm d}t^i/\dt = i t^{i-1}$, and we have
\begin{equation}
\Qvis{i}{m} = \int_0^1 i t^{i-1} \Qvis{i}{m}(t) \,\dt \enspace .
\label{eq:visdegt}
\end{equation}
Our goal is then to calculate $\Qvis{i}{m}(t)$.

Let us start by calculating the probability $\Pvis[t]$ that, if $v$ 
has index $t$, a given copy of $v$ other than the copy 
with index $t$---that is, a given copy which is added to the
queue at time $t$---gives rise to a visible edge.  
Call this copy $u$, and call its partner $w$. 
According to Algorithm \ref{alg:BFS}, the edge $(u,w)$ is visible if and
only if (1) $u$ makes it to the head of the queue without being matched
first, and (2) when it does, $w$ is still untouched.  But (1) is
equivalent to saying that $w$ is untouched at time $t$, since if 
$w$ is already in the queue at time $t$, it is ahead of $u$, 
and $u$ will be matched before it reaches the head of the queue.  
Similarly, (2) is equivalent to saying that {\em all of $w$'s siblings'
partners} are untouched at time $t$, since if any of these are already 
in the queue at time $t$, and thus ahead of $u$, then 
$w$'s vertex will be touched, and $w$ enqueued, 
by the time $u$ reaches the head of the queue.

Given the number of untouched and unexposed copies 
$\Cunto[t]$ and $\Cunex[t]$ at the time $t$ when $u$ joins the queue, 
the probability that its uniformly random partner $w$ is
untouched is $\PUNTO[t] = \Cunto[t] / \Cunex[t]$.  
Conditioning on this event, the probability that $w$ belongs to a
vertex with degree $k$ is 
$\Punto[t]{k} = k \Vunto[t]{k} / \Cunto[t]$.
We require that the partners of $w$'s $k-1$ siblings are
also untouched.  If we ignore the fact that we are choosing untouched copies
without replacement (and that one untouched copy has already been taken 
for $w$), and if we assume that $v$, its neighbors, and its neighbors' neighbors 
form a tree (\ie that $v$ does not occur in a triangle or 4-cycle, and that neither it
nor its neighbors have any multiple edges), then the probability that these $k-1$ copies 
are all untouched is $\PUNTO[t]^{k-1}$.  This gives
\begin{align} 
\label{eq:bigpvis}
\Pvis[t] 
& =  \PUNTO[t] \sum_k \Punto[t]{k} \,\PUNTO[t]^{k-1}  \nonumber\\
& =  \sum_k \Punto[t]{k} \,\PUNTO[t]^k .
\end{align}

Since \Vunto[t]{k}, \Cunto[t] and \Cunex[t] are concentrated according
to Lemma~\ref{lem:copies}, substituting their expectations then
gives a good approximation for \Pvis[t], namely
\begin{equation}
\pvis[t] 
= \sum_k \frac{k a_k t^k}{\cunto[t]} \left( \frac{\cunto[t]}{\cunex[t]} \right)^{\!k} \enspace . 
\label{eq:pvis}
\end{equation}
Then, if we neglect the possibility
of self-loops and parallel edges involving $u$ and its siblings, 
and again ignore the fact that we are choosing without replacement 
(\ie that processing each sibling changes $\Cunto$, $\Cunex$, and $\Pvis$ slightly) 
the events that each of $u$'s siblings give rise to a visible edge are
independent, and the number $m$ of visible edges is approximately
binomially distributed as  $\bin(i-1,\pvis[t])$.  

We wish to confirm this analysis by showing that 
\whp\ $v$, its neighbors, and its neighbors' neighbors form a tree.
It is easy to show this for graphs with bounded degree; however, for power-law degree 
distributions $a_k \sim k^{-\alpha}$, it is somewhat
delicate, especially for $\alpha$ close to $2$.  
The following lemmas show that there are very few vertices of very high degree, 
and then show that the above is \whp\ true of $v$ if $v$ has sufficiently low degree.  
We then show that we can think of all the copies involved as chosen with replacement.
Recall that the mean degree $\delta = \sum_j j a_j$ is finite, and
let $\beta < \min(\frac{1}{2},\frac{(\alpha-2)}{2})$ as in Lemma~\ref{lem:copies}.

\begin{lemma} 
\label{lem:nbrs}
The probability that a random copy belongs to a vertex of degree
greater than $k$ is $o(k^{-2\beta})$.
\end{lemma}

\begin{proof}  This probability is
\[ 
\frac{\sum_{j > k} j a_j}{\sum_j j a_j} 
< \frac{C}{\delta} \sum_{j > k} j^{1-\alpha} 
< \frac{C}{\delta(2-\alpha)} k^{-(\alpha-2)} = o(k^{-2\beta}) \enspace . 
\]
\end{proof}

\begin{lemma}
\label{lem:nbrhood}
There are constants $\gamma > \eta > 0$ such that if $v$ is a vertex
of degree $i < n^\eta$, then the probability that $v$ or its neighbors
have a self-loop or multiple edge, or that $v$ is part of a triangle
or a cycle of length $4$, is $o(n^{-\gamma})$.  Thus, $v$, its neighbors, and 
its neighbors' neighbors form a tree with probability $1-o(n^{-\gamma})$.
\end{lemma}

\begin{proof}
First, we employ Lemma~\ref{lem:nbrs} to condition on the event that
none of $v$'s neighbors have degree greater than $n^\lambda$, where
$\lambda$ (and $\eta$) will be determined below.  By a union bound
over these $i < n^\eta$ neighbors, this holds with probability
$1-o(n^{\eta-2\lambda\beta})$.  (Unfortunately, we cannot also
condition on $v$'s neighbors' neighbors having degree at most
$n^\lambda$ without breaking this union bound.) 

Now, if we choose two copies independently and uniformly at random,
the probability that they are both copies of a given vertex of degree
$j < n^\lambda$ is $j(j-1)/(\delta n)^2 < n^{2\lambda-2}$, and the
probability that they are both copies of {\em any} such vertex is at
most $n^{2\lambda-1}$.  Moreover, the probability that two random
copies are siblings, regardless of the degree of their vertex, is
\[ \Psib = \frac{\sum_j j(j-1) a_j n}{\left( \sum_j j a_j n \right)^2} 
< \frac{1}{\delta^2 n} \sum_j j^2 a_j = o(n^{-2\beta}) \enspace . \]

Taking a union bound over all pairs of copies of $v$, the probability
that $v$ has a multiple edge, \ie that two of its copies are matched
to copies of the same neighboring vertex, is at most 
$i^2 n^{2\lambda-1} = O(n^{2\eta+2\lambda-1})$, and the
probability that $v$ contains a self-loop, \ie that two of its copies
are matched, is $O(i^2/(\delta n)) = O(n^{2\eta-1})$.  For each of
$v$'s neighbors, the probability of parallel edges involving it is
at most $n^{2\lambda} \Psib = o(n^{2\lambda-2\beta})$, and the
probability of a self-loop is $O(n^{2\lambda}/(\delta n)) = O(n^{2\lambda-1})$.
Taking a union bound over all of $v$'s neighbors, the probability that
any of them have a self-loop or multiple edge is
$o(n^{\eta+2\lambda-2\beta})$.

To determine the expected number of triangles containing $v$, we
notice that any such triangle contains two copies each from $v$ and
two of its neighbors, and edges between the appropriate pairs. 
A given pair of copies is connected with probability 
$O(1/(\delta n))$, so the expected number is 
\begin{align*}
 O\!\left( n^2 \,n^{2\eta} (n^{2\lambda})^2/(\delta n)^3 \right) 
& =  O( n^{2\eta+4\lambda-1} ) \enspace .
\end{align*}

Similarly, each 4-cycle involves two copies each of $v$ and two of its
neighbors, such that one copy from each of the neighbors is matched
with one copy of $v$, and the other two copies are matched with copies
of the same node. Thus, the expected number of 4-cycles involving $v$ is
\begin{align*}
O\!\left( n^2 \,n^{2\eta} (n^{2\lambda})^2 \Psib /(\delta n)^2\right) 
& =  o(n^{2\eta + 4\lambda - 2\beta}) \enspace .
\end{align*}

Collecting all these events, the probability that the statement of the
lemma is violated is 
\[ o(n^{-\gamma}) \mbox{ where } \gamma = -\max(
\eta-2\lambda\beta, 
2\eta + 4\lambda - 2\beta ) 
\enspace . \]
If we set $\eta = \beta^2/6$ and $\lambda = \beta/4$, then $\gamma = \beta^2/3$.
\end{proof}

The next lemma shows that, conditioning on the event of
Lemma~\ref{lem:nbrhood}, the copies discussed in our analysis above
can be thought of as chosen with replacement, as long as we are not 
too close to the end of the process where untouched copies become rare.  
Therefore, the number of visible edges is binomially distributed.

\begin{lemma}
\label{lem:binom}
Let $\eta,\gamma$ be defined as in Lemma~\ref{lem:nbrhood}.  There
exists a constant $\theta > 0$ such that for $t \in [n^{-\theta},1]$
and $i < n^{-\eta}$, 
\begin{align*}
\Abs{\Qvis{i}{m}(t) - \Prob{\bin(i-1,\Pvis(t)) = m}}
& <  n^{-\gamma} \enspace .
\end{align*}
\end{lemma}

\begin{proof}
Let $\dmin$ be the minimum degree of the graph, \ie the smallest $j$
such that $a_j > 0$.  Note that $\dmin \ge 3$, and set 
$\theta = \beta/(2 \dmin) < 1/12$.  
For $t \ge n^{-\theta}$, 
we have that $\Expect{\Cunex[t]}=\delta t^2 n=\Omega(n^{1-\beta/\dmin})$,
and this bound holds \wohp\ by Lemma~\ref{lem:copies}.

Conditioning on $v$'s neighbors having degree at most $n^\lambda$ as in 
Lemma~\ref{lem:nbrhood}, the number of visible edges of $v$ is
determined by a total of at most $n^{\eta+\lambda}$ copies.  These are
chosen without replacement from the unexposed copies. If we instead
choose them with replacement, the probability of a collision in which
some copy is chosen twice is at most 
$(n^{\eta+\lambda})^2 / \Cunex[t] = O(n^{2\eta+2\lambda+\beta/\dmin-1})=o(n^{-1/2})$.
This can be absorbed into the probability $o(n^{-\gamma})$ that the
statement of Lemma~\ref{lem:nbrhood} does not hold.  If there are no
collisions, then we can assume the copies are chosen with replacement,
and each of $v$'s $i-1$ outgoing edges is independently visible with
probability $\Pvis[t]$, as defined in~\eqref{eq:bigpvis}. 
\end{proof}

The next three lemmas then show that
$\pvis[t]$ is a very good approximation for $\Pvis[t]$ for most $t$, and that therefore
the distribution of $\Qvis{i}{m}(t)$
is very close to $\bin(i-1,\pvis[t])$.

\begin{lemma}
\label{lem:pvis}  
Let $\gamma$ and $\theta$ be defined as in Lemma~\ref{lem:nbrhood} and
Lemma~\ref{lem:binom}.  There exists a constant $\kappa > 0$ such that
for all $t \in [n^{-\theta},1-n^{-\kappa}]$, \wohp\ $\Abs{\Pvis[t] - \pvis[t]} < n^{-\gamma}$.
\end{lemma}

\begin{proof} 
Recall that $\theta = \beta/(2\dmin)$.  From Lemma~\ref{lem:copies}, since $t \ge n^{-\theta}$
we have \wohp\ $\Cunex[t]=\Omega(t^2 n) = \Omega(n^{1-\beta/\dmin})$ as in the
previous lemma, and $\Cunto[t] = \Omega(t^\dmin n) =
\Omega(n^{1-\beta/2})$.  For definiteness, take $\eps = 1/12$ in 
Lemma~\ref{lem:copies}; then \wohp
\begin{align*}
\Cunex[t] & =  \cunex[t] n + o(n^{7/12}) 
 =  \cunex[t] n \cdot (1+o(n^{\beta/\dmin-5/12})) \\ 
\Cunto[t] & =  \cunto[t] n + o(n^{1-\beta}) 
 =  \cunto[t] n \cdot (1+o(n^{-\beta/2})) \\ 
\Vunto[t]{k} & =  a_k t^k n + o(n^{7/12}) \enspace . 
\end{align*}
Recall that $\beta < 1/2$ and $\dmin \ge 3$.  Since $\beta/\dmin - 5/12 < -1/4 < -\beta/2$, 
\[ \begin{array}{lclcl}
\PUNTO[t] & = & \displaystyle{\frac{\Cunto[t]}{\Cunex[t]} }
& = & \displaystyle{\frac{\cunto[t]}{\cunex[t]} \left( 1 + o(n^{-\beta/2}) \right)} \enspace .
\end{array} \]
Now, we compare \Pvis[t] with \pvis[t] term by term, and separate their
respective sums into the terms with $3 \le k \le n^{\beta/12}$ and
those with $k > n^{\beta/12}$.  For all $k \le n^{\beta/12}$, 
since $\beta/12 + \beta/2 - 5/12 < -1/8 < -\beta/4$,
\[ \begin{array}{lclcl}
\Punto[t]{k} & = & \displaystyle{\frac{k \Vunto[t]{k}}{\Cunto[t]}}
& = & \displaystyle{ \frac{k a_k t^k}{\cunto[t]} + o(n^{-\beta/4}) } \enspace , 
\end{array} \]
and since $(1+x)^y = 1+O(xy)$ if $xy < 1$,
\begin{align*}
\PUNTO[t]^k 
& = \left( \frac{\cunto[t]}{\cunex[t]} \right)^{\!k} \left( 1 + o(n^{-\beta/2}) \right)^k \\
& = \left( \frac{\cunto[t]}{\cunex[t]} \right)^{\!k} \left( 1 + o(n^{-5\beta/12}) \right) 
\enspace .
\end{align*}
Thus, each term obeys
\Equation{\Punto[t]{k} \PUNTO[t]^k}{\frac{k a_k t^k}{\cunto[t]} \left( \frac{\cunto[t]}{\cunex[t]} \right)^{\!k}
+ o(n^{-\beta/4}) }
and the total error from the first $n^{\beta/12}$ terms is
at most $n^{\beta/12} \cdot o(n^{-\beta/4}) = o(n^{-\beta/6})$.

On the other hand, if $k > n^{\beta/12}$, then for any 
$t \le 1-n^{-\kappa}$ we have  
\[ \begin{array}{lclcl}
t^k & < & \displaystyle{\e^{-kn^{-\kappa}}} & < & \displaystyle{\e^{-n^{\beta/12-\kappa}}} \enspace ,
\end{array} \]
Setting $\kappa < \beta/12$ makes this exponentially small.  
In that case, taking a union bound 
over all $n^{\beta/12} < k < n$, \wohp\ there are no unexposed
vertices of degree greater than $n^{\beta/12}$; thus $\Punto[t]{k} = 0$
and these terms of \Pvis[t] are zero.  The corresponding terms of
\pvis[t] are exponentially small as well, so the total error from these terms
is exponentially small.  Thus, the total error is $o(n^{-\beta/6})$, and 
since $\gamma = \beta^2/3 < \beta/6$, this can be absorbed into the 
probability $o(n^{-\gamma})$ that the conditioning of Lemma~\ref{lem:nbrhood}
is violated.
\end{proof}


\begin{lemma}  
\label{lem:bin}
For any $s, m, p$ and $\Delta$, 
\LEquation{\Abs{\! \Prob{\bin(s,p)=m} - \Prob{\bin(s,p+\Delta)=m}}}{s \Delta \enspace.}
\end{lemma}

\begin{proof}
It is sufficient to bound the derivative of
these probabilities with respect to $p$ as follows.
\begin{align*}
\Abs{\frac{\p}{\p p} \Prob{\bin(s,p)=m}} 
& =  \Abs{\frac{\p}{\p p} {s \choose m} p^m (1-p)^{s-m}} \\
& =  {s \choose m} p^m (1-p)^{s-m} \Abs{\frac{m}{p} - \frac{s-m}{1-p}} \\
& \leq  {s \choose m} p^m (1-p)^{s-m} \,\max\left( \frac{m}{p} , \frac{s-m}{1-p} \right) \\
& \leq  \max\left(  
\sum_{m=0}^s {s \choose m} \frac{m}{p} \,p^m (1-p)^{s-m} \right.\; , \; 
\left.\sum_{m=0}^s {s \choose m} \frac{s-m}{1-p} \,p^m (1-p)^{s-m} 
\right) \\
& =  \max( s, s ) \; = \; s \enspace . 
\end{align*}
\end{proof}

\begin{lemma} \label{lem:qvis}
Let \SET{a_j} be a \reasonable degree distribution and assume that
$G$ is connected.  There are
constants $\theta, \kappa,$ $\eta, \mu > 0$, such that for all 
$t \in [n^{-\theta},1-n^{-\kappa}]$ and all $i < n^{\eta}$,  
for sufficiently large $n$,
\begin{align*}
\Abs{\Qvis{i}{m}(t) - \Prob{\bin(i-1, \pvis[t]) = m}} 
& <  n^{-\mu} \enspace ,
\end{align*}
where $\pvis[t]$ is defined in~\eqref{eq:pvis}.
\end{lemma}

\begin{proof}
Given Lemma~\ref{lem:pvis}, we apply Lemma~\ref{lem:bin} and the triangle inequality.  
In this case, we have $s \leq n^\eta$ and $\Delta < n^{-\gamma}$, so the
error in $\Qvis{i}{m}$ is at most $n^{\eta-\gamma}$.  Recalling from the
proof of Lemma~\ref{lem:nbrhood} that  
$\eta = \beta^2/6$ and $\gamma = \beta^2/3$, for sufficiently large $n$ this is 
less than $n^{-\mu}$ for any $\mu < \beta^2/6$.  
\end{proof}

Finally, combining Lemma~\ref{lem:qvis} with~\eqref{eq:visdeg},~\eqref{eq:visdegt}, 
and~\eqref{eq:pvis}, if
\begin{align}
a^\obs_{m+1}
& = \sum_i a_i \Bigg[ \int_0^1 i t^{i-1} \,{i-1 \choose m} \,\pvis[t]^m
\,(1-\pvis[t])^{i-1-m} \, \dt \Bigg] \enspace,
\label{eq:aobs} 
\end{align}
where, combining~\eqref{eq:pvis} with~\eqref{eq:unto} and~\eqref{eq:unex}, 
\begin{align*}
\pvis[t]
& = \frac{1}{\sum_j j a_j t^j} \sum_k k a_k t^k \left( 
\frac{\sum_j j a_j t^j}{\delta t^2} \right)^{\!k} \enspace ,
\end{align*}
then we have the following lemma. 
\begin{lemma}
\label{lem:exp}
Let $\SET{a_i}$ be a \reasonable degree sequence and assume that $G$
is connected.  There is a constant $\zeta > 0$ such that for sufficiently large $n$, 
for all $j < n$ 
\begin{align*}
\Abs{\Expect{\VisDeg{j}} - a^\obs_j n} & <  n^{1-\zeta} \enspace . 
\end{align*}
\end{lemma}
 
\begin{proof}
There are three sources of error in our estimate of $\Expect{\VisDeg{j}}$ for each $j$.  
These are the error $n^{-\mu}$ in $\Qvis{i}{m}(t)$ given by Lemma~\ref{lem:qvis}, 
and the fact that two types of vertices are not covered by that lemma: those with 
degree greater than $n^\eta$, and those which join the queue at some time
$t \notin [n^{-\theta},1-n^{-\kappa}]$.  The total error is then at most $n^{1-\mu}$ 
plus the number of vertices of either of these types.
The number of vertices of degree greater than $n^\eta$ is at most
\[ n \sum_{j > n^\eta} a_j 
\;\;<\;\; C n \sum_{j > n^\eta} j^{-\alpha}
\;\;=\;\; O(n^{1-(\alpha-1)\eta})
\enspace .
\]
The number of vertices that join the queue at a time 
$t \notin [n^{-\theta}, 1-n^{-\kappa}]$  
is at most the number of copies whose index is outside this interval.  This is
binomially distributed with mean $n^{1-\theta}+n^{1-\kappa}$, 
and by the Chernoff bound, this is \wohp\ less than $n^{1-\zeta}$ for 
sufficiently large $n$ for any $\zeta < \min(\theta,\kappa)$.  
The (exponentially small) probability that this bound is violated 
can be absorbed into $n^{1-\zeta}$ as well.
Setting $\zeta < \min(\mu,(\alpha-1)\eta,\theta,\kappa)$ completes the proof.
\end{proof}

\section{Concentration}
\label{sec:concentration} 

In this section, we prove that the number \VisDeg{j} of nodes of observed 
degree $j$ is tightly concentrated around its expectation \Expect{\VisDeg{j}}. 
Specifically, we prove
\begin{theorem} \label{thm:concentration}
There is a constant $\rho > 0$ such that,  
with overwhelmingly high probability, the following holds
simultaneously for all $j$:
\LEquation{\Abs{\VisDeg{j}-\Expect{\VisDeg{j}}}}{O(n^{1-\rho}) \enspace .}
\end{theorem}

\begin{proof}
In order to prove concentration, the style of analysis in the
previous section will not be sufficient. Intuitively, the reason
is that changing a single edge in the graph can have a dramatic
impact on the resulting BFS tree, and thus on the observed
degree of a large number of vertices. As a result, it seems unlikely
that \VisDeg{j} can be decomposed into a large number of
small contributions such that their sum can easily be shown to be
concentrated. In particular, this rules out the 
direct application both of Chernoff-style bounds and of
martingale-based inequalities.

There is, however, a sense in which martingale bounds will prove
helpful. The key is to decompose the evolution of \VisDeg{j} into a
small number of ``bulk moves,'' and prove 
concentration for each one of them. Concretely, assume that the
BFS tree has already been exposed up to a certain distance $r$ from
the root, and that we know the number of copies in the queue, 
as well as the number of untouched copies at that point. Since all
these copies will be matched uniformly at random, one can use an
edge-switching martingale bound to prove that the degree distributions
of nodes at distance $r+1$ from the root will be sharply
concentrated. In fact, this concentration argument applies to the
observed degrees of the neighbors of any ``batch'' of copies that comprise the
queue \QueueSet[t] at some time $t$.

We will implicitly divide the copies in the graph into such
``batches'' by specifying a set of {\em a priori fixed} points in time
at which we examine the system. That is, we will approximate
\VisDeg{j} by the sum of the observed degrees of the neighbors of
\QueueSet[t] over these time steps. We will show that each of the terms
in the sum is sharply concentrated around its expectation, and then
prove that the true expectation of \VisDeg{j} is not very far from the
expectation of the sum that we consider. For the latter part, it is
crucial that most vertices be counted exactly once in the sum; this
will follow readily from the concentration given by Lemma~\ref{lem:copies}.

To make the above outline precise, we let 
$\Queue[t] := \SetCard{\QueueSet[t]}$ be the number of copies in the
queue at time $t$, and let 
\[ \queue[t] := \Expect{\Queue[t]} = (\cunex[t] - \cunto[t]) \cdot n \] 
be its expected size. We define a sequence of $r \leq \log^2 n$
times at which we observe the queue and its neighbors.
We start with $t_1 = 1$. For each $i$, we let $t_{i+1} \geq 0$ be
maximal such that 
\[ \cunex[t_i] - \cunex[t_{i+1}] \geq \queue[t_i]/n + 2n^{-\beta} \enspace , \] 
where $\beta$ is defined as in Lemma~\ref{lem:copies}.  
Depending (deterministically) on the properties of the real-valued functions 
$\cunex$ and $\cunto$, there may be an $i < \log^2 n$ such
that $t_{i+1}$ does not exist, namely when 
$\cunex[t_i] < 2n^{-\beta}$. If so, we let $r$ be that
$i$; otherwise, we let $r = \log^2 n$.

For each degree $j$, let \VisDeg[i]{j} denote the number of
vertices adjacent to \QueueSet[t_i] whose observed degree is
$j$. Lemma~\ref{lem:batch-concentration} below shows that
each \VisDeg[i]{j} is sharply concentrated. However, we want to
prove concentration for the overall quantity \VisDeg{j}. Using a union bound 
over all $i=1,\ldots,r$ and summing up the corresponding
\VisDeg[i]{j} will give us concentration for \VisDeg{j}, assuming that
(1) not too many times $t_i$ are considered, (2) nodes are not
double-counted for multiple $i$, and (3) almost all nodes are
considered in some batch $i$.  

For the first point, recall that we explicitly chose $r = O(\log^2 n)$.
For the second, observe that whenever 
\[ \Cunex[t_i] - \Cunex[t_{i+1}] \;\geq\; \queue[t_i] + 2n^{1-\beta} \;\geq\; \Queue[t_i] \]
for all times $i$, then all of the  \QueueSet[t_i] are disjoint.
Each of these bounds holds \wohp\ by Lemma~\ref{lem:copies},
and by the union bound, \wohp\ they hold simultaneously.

This leaves the third point. Here, we first bound the number of
nodes that remain unexposed after time $t_r$. If the construction
terminated prematurely (\ie $r < \log^2 n$), then the fact that
$\cunex[0] = 0$ implies that $\cunto[t_r] < 2n^{-\beta}$, so by 
Lemma~\ref{lem:copies}, at most
$O(n^{1-\beta})$ copies remain unexposed \wohp
On the other hand, when $r = \log^2 n$, we can use the
fact that the diameter of a random graph is bounded by 
$\log^2 n$ with probability at least $1-n^{-1/2}$, which we prove
in the full paper using techniques of Bollob\'{a}s and Chung
\cite{bollobas:chung:cycle+matching}. 
Even if \Cunto[t_r] were $\Omega(n)$ in the remaining case, since this 
occurs with probability at most $n^{-1/2}$, we have 
$\cunto[t_r] \cdot n = O(n^{1/2}) = O(n^{1-\beta})$ since $\beta < 1/2$.

Let \Event{E} denote the event that $\Abs{\Vunto[t_i]{j} -
\vunto[t_i]{j}} \leq n^{1/2+\epsilon}$ and $\Abs{\Queue[t_i] -
\queue[t_i]} \leq 2n^{1-\beta}$ hold simultaneously for all $i$.
By Lemma~\ref{lem:copies}, \Event{E} occurs \wohp
In that case, we know that (1) all of the sets \QueueSet[t_i] are disjoint, and (2)
the union of all the \QueueSet[t_i] excludes at most $2r \cdot
n^{1-\beta} + O(n^{1-\beta}) = \tilde{O}(n^{1-\beta})$ copies
total (where $\tilde{O}$ includes ${\rm polylog}(n)$ factors). 
Thus, $\Abs{\VisDeg{j} - \sum_{i=1}^r \VisDeg[i]{j}} = 
\tilde{O}(n^{1-\beta})$ \wohp, which implies that
$\Abs{\Expect{\VisDeg{j}} -\sum_{i=1}^r \Expect{\VisDeg[i]{j}}}
= \tilde{O}(n^{1-\beta})$, since this difference is
deterministically bounded above by $n$.

By Lemma~\ref{lem:batch-concentration} below and a union bound
over all $i$, there is a $\tau > 0$ such that \wohp\ $\Abs{\VisDeg[i]{j} -
\Expect{\VisDeg[i]{j}}} = O(n^{1-\tau})$ holds simultaneously 
for all $i=1,\ldots,r$ and all $j$. Hence, by a union bound with
the event \Event{E}, and the triangle inequality, the following holds
\wohp: 
\begin{align*}
\Abs{\VisDeg{j} - \Expect{\VisDeg{j}}} &
 \leq \Abs{\VisDeg{j} - \sum_{i=1}^r \VisDeg[i]{j}} 
 \;+ \sum_{i=1}^r \Abs{\VisDeg[i]{j} - \Expect{\VisDeg[i]{j}}} 
 \;+ \Abs{\Expect{\VisDeg{j}} -\sum_{i=1}^r \Expect{\VisDeg[i]{j}}} \\
& = \tilde{O}(n^{1-\beta}) + \tilde{O}(n^{1-\tau}) \\ 
& = O(n^{1-\rho}) \enspace .
\end{align*}
for any $\rho < \min(\beta,\tau)$, completing the proof of Theorem \ref{thm:concentration}.
\end{proof}

The concentration for one ``batch'' of nodes at time $t_i$ is captured
by the following lemma.

\begin{lemma} \label{lem:batch-concentration}  There is a constant $\tau > 0$
such that, for any fixed $i$, \wohp, $\Abs{\VisDeg[i]{j} - \Expect{\VisDeg[i]{j}}}
= O(n^{1-\tau})$ holds simultaneously for all $j$.
\end{lemma}

\begin{proof}
As explained above, the idea for the proof is to apply an
edge-exposure Martingale-style argument to the nodes that are adjacent
to \QueueSet[t_i]. We use the following concentration
inequality for random variables on matchings due to 
Wormald~\cite[Theorem~2.19]{wormald:random-graph-models}.  
A {\em switching} consists of replacing two edges
\SET{p_1,p_2}, \SET{p_3,p_4} by \SET{p_1,p_3}, \SET{p_2,p_4}.

\begin{theorem} \cite{wormald:random-graph-models}
\label{thm:wormald-martingale}
Let $X_k$ be a random variable defined on uniformly random
configurations $M,M'$ of $k$ copies, such that, whenever $M$ and $M'$
differ by only one switching, 
\[ \Abs{X_k(M) - X_k(M')} \leq c \] 
for some constant $c$. Then, for any $r > 0$,
\begin{align*}
\Prob{\Abs{X_k - \Expect{X_k}} \geq \Delta} & <  2 \e^{-\Delta^2/(kc^2)} \enspace .
\end{align*}
\end{theorem}

For fixed values $q$ and $\bvec=b_1,\ldots, b_n$, let
\Event[q,\bvec]{E} denote the event that $\Queue[t_i] = q \mbox{
and } \Vunto[t_i]{j} = b_j \mbox{ for all } j$. Conditioned on
\Event[q,\bvec]{E}, the matching on the $q+\sum_j j b_j$ copies is
uniformly random.  Since any switching changes the
value of \VisDeg[i]{j} by at most 2, Theorem \ref{thm:wormald-martingale} 
implies that 
\begin{equation}
\label{eq:bj}
\Abs{\VisDeg[i]{j} - \ExpectC{\VisDeg[i]{j}}{\Event[q,\bvec]{E}}}
\le n^{1/2+\eps}
\end{equation} 
holds \wohp\ for any $\eps > 0$.
If we knew the queue size $q$ and the number $b_j$ of
untouched nodes of degree $j$ exactly, then we could apply
Theorem~\ref{thm:wormald-martingale} directly. 

In reality, we will certainly not know the precise values of $q$ and
\bvec.  Therefore, we need to analyze the effect that deviations of these
quantities will have on our tail bounds.
We do this by showing in
Lemma~\ref{lem:visdeg-expectations} below that the conditional
expectations \ExpectC{\VisDeg[i]{j}}{\Event[q,\bvec]{E}} 
are close to the actual expectations \Expect{\VisDeg[i]{j}}.  
It follows that concentration around the conditional expectation 
implies concentration around the actual expectation.
Specifically, write
\[ I^q := \left[ \queue[t_i] - 2n^{1-\beta}, \queue[t_i] + 2n^{1-\beta} \right] \]
for the interval of possible queue lengths under consideration, and, 
for some $0 < \eps < 1/2$, write
\[ I^b_j := \left[ \vunto[t_i]{j} - n^{1/2+\epsilon}, \vunto[t_i]{j} + n^{1/2+\epsilon} \right] \]
for the interval of possible numbers of untouched vertices of degree $j$,
as well as $I^b := I^b_1 \times \cdots \times I^b_n$ for the range
of all possible combinations of numbers of untouched vertices.
Now, let \Event[\leq]{E} be the event that
$\Queue[t_i] \in I^q$ and $\Vunto[t_i]{j} \in I^b_j$ for all $j$.
Notice that \Event[\leq]{E} occurs \wohp\ by Lemma~\ref{lem:copies}.

Lemma \ref{lem:visdeg-expectations} ensures that whenever
$q \in I^q$ and $\bvec \in I^b$, then 
the conditional expectation is close to the true expectation, \ie 
for some $\tau > 0$, 
\[ \Abs{\ExpectC{\VisDeg[i]{j}}{\Event[q,\bvec]{E}} - \Expect{\VisDeg[i]{j}}} 
= O(n^{1-\tau}) \enspace . \] 
Thus, for all such $q$ and $\bvec$, combining this with~\eqref{eq:bj}
and the triangle inequality gives
$\Abs{\VisDeg[i]{j} - \Expect{\VisDeg[i]{j}}} = O(n^{1-\tau})$, 
so the latter occurs \wohp
Finally, a union bound with the event \Event[\leq]{E} and over all
$j$ completes the proof.
\end{proof}

The final missing step is a bound relating the conditional expectation
of \VisDeg[i]{j} with its true expectation. Intuitively, since all
relevant parameters are sharply concentrated, one would expect that
the conditional expectation for any of the likely values is close to
the true expectation. Making this notion precise turns out to be
surprisingly cumbersome.

\begin{lemma} \label{lem:visdeg-expectations} 
There is a constant $\tau > 0$ such that, 
for any $q \in I^q$ and $\bvec \in I^b$, we have
\[ \Abs{\ExpectC{\VisDeg[i]{j}}{\Event[q,\bvec]{E}} - \Expect{\VisDeg[i]{j}}}
= O(n^{1-\tau}) \enspace . \]
\end{lemma}

\begin{proof}
We first compare the conditional expectations for two ``scenarios'' of
queue lengths and untouched vertices when the scenarios are
close. We will see that the conditional expectations in those two
scenarios will be close; from that, we can then conclude that any
conditional expectation is close to the true expectation.

Given $q, q'$ and $\bvec, \bvec'$,
such that $\Abs{q-q'} \leq 4n^{1-\beta}$, and
$\Abs{b_j-b'_j} \leq 2n^{1/2+\epsilon}$ for each $j$,
we let $\hat{q} = \min(q,q')$ and
$\hat{b}_j = \min(b_j,b'_j)$, and define the events 
$\Event{E} := \Event[q,\bvec]{E}, \Event{E'} := \Event[q',\bvec']{E}$, and
$\hat{\Event{E}} := \Event[\hat{q}, \hat{\bvec}]{E}$.
Now, we claim that, for some $\tau > 0$, 
\Equation{\Abs{\ExpectC{\Vunto[t_i]{j}}{\Event{E}}
- \ExpectC{\Vunto[t_i]{j}}{\hat{\Event{E}}}}}{O(n^{1-\tau})}
for all $j$, and similarly for \Event{E'}. By the triangle inequality,
this immediately implies that
\Equation{\Abs{\ExpectC{\Vunto[t_i]{j}}{\Event{E}}
- \ExpectC{\Vunto[t_i]{j}}{\Event{E'}}}}{O(n^{1-\tau}) \enspace .}

To prove the claim, imagine that in the $(q,\bvec)$ instance, 
we color an arbitrary, but fixed, set of $q-\hat{q}$ of copies in the queue
black, as well as the copies of an arbitrary set of
$b_j-\hat{b}_j$ vertices for each degree $j$.
To expose the matching, we first expose all the neighbors of black
copies, and color them blue, and then choose a uniform matching among
the remaining (white, say) copies.
The number of blue copies obeys some distribution
$D_{q,\bvec}$, but in any case, it never exceeds the
total number of black copies.  Since $q,q' \in I^q$ and $\bvec,\bvec' \in I^b$, 
for any $\nu > 0$, this total number is at most
\begin{align*}
(q-\hat{q}) + \sum_j j \cdot (b_j-\hat{b}_j) 
& \leq  4n^{1-\beta} + 2 n^{1/2+\epsilon} \sum_{j \le n^\nu} j
 + 2 \sum_{j > n^\nu} j \cdot a_j n\\
& =  4n^{1-\beta} + O(n^{1/2+\eps+2\nu}) + O(n^{1-(\alpha-2)\nu}) \\
& = O(n^{1-\tau}) \enspace ,
\end{align*}
for any $\tau < \min(\beta,1/2-\eps-2\nu,(\alpha-2)\nu)$.  Note that $\tau > 0$
as long as $1/2-\eps-2\nu > 0$; recall that we took $\eps < 1/2$ in the previous lemma, 
so we can choose any $\nu < (1/2-\eps)/2$.

Now, in the $(\hat{q},\hat{\bvec})$ instance, we
can generate a uniformly random matching as follows: 
we choose a number $k$ according to $D_{q,\bvec}$, choose $k$ copies
uniformly at random and color them blue, and determine a uniformly
random matching among the blue copies only. Then, we match up the
remaining white copies uniformly at random. We will call
a node black if at least one of its copies is black, blue if at least
one of its copies is blue, and white otherwise.

Since the set of nodes that are not black is deterministically the
same in both instances, and the probability distribution of blue nodes
is the same in both, the expected number of white nodes that end up
with visible degree $j$ is the same in both experiments. Hence, the
expected total number of nodes with observed degree $j$ can only differ
by the number of blue or black nodes. Even if the degrees of those
nodes were chosen adversarially, the difference cannot be more than
$O(n^{1-\tau})$, since this is a
deterministic upper bound on the number of black or blue copies, 
and hence on the number of black or blue nodes.
By summing up over the entire probability space, this now proves the
claim for \Event{E} and $\hat{\Event{E}}$, and thus also for
\Event{E} and \Event{E'}.

We know that if $q,q' \in I^q$ and $\bvec, \bvec' \in I^b$, then they
always satisfy the necessary conditions, and hence the conditional
expectations are within $O(n^{1-\tau})$. Summing up over all 
$q \in I^q$ and $\bvec \in I^b$ therefore shows that
\begin{align*}
\Abs{\ExpectC{\VisDeg[i]{j}}{\Event[q,\bvec]{E}} - \ExpectC{\VisDeg[i]{j}}{\Event[\leq]{E}}} = O(n^{1-\tau}) \enspace .
\end{align*}
Finally, because \Event[\leq]{E} occurs with overwhelmingly
high probability, and \VisDeg[i]{j} is bounded by $n$, we obtain that, for all $c$, 
\begin{align*}
\Abs{\ExpectC{\VisDeg[i]{j}}{\Event[\leq]{E}} - \Expect{\VisDeg[i]{j}}}
= O(n^{-c}) \enspace ,
\end{align*}
and the triangle inequality completes the proof.
\end{proof}

\section{Generating functions}
\label{sec:genfunc}

In this section, we use the formalism of generating
functions~\cite{wilf:generatingfunctionology} to express the results
of Section~\ref{sec:expected} more succinctly, and complete the proof of
Theorem~\ref{thm:genfunc}.  Given the generating function of the
degree sequence of the underlying graph
\Equation{g(z)}{\sum_i a_i z^i \enspace ,}
our goal is to obtain the generating function for the expected degree sequence
of the breadth-first tree as approximated by Lemma~\ref{lem:exp},
\Equation{g^{\obs}(z)}{\sum_i a^\obs_i z^i \enspace .}
Using the generating function formalism, we can write 
\[ \cunto[t] = t g'(t) , \; 
\delta = g'(1) , \; 
\cunex[t] = t^2 g'(1) \enspace , \] 
and from~\eqref{eq:pvis} we have
\begin{align}
\pvis[t]
& = \sum_{k} \frac{k a_k t^k}{t g'(t)}\left(\frac{g'(t)}{tg'(1)}\right)^k \nonumber \\
& = \frac{1}{tg'(t)} \sum_k ka_{k} \left(\frac{g'(t)}{g'(1)}\right)^k \nonumber \\
& = \frac{1}{tg'(1)} \,g'\!\left( \frac{g'(t)}{g'(1)}\right) \enspace . 
\label{eq:pvisgen}
\end{align}
Then, combining~\eqref{eq:pvis} and \eqref{eq:aobs}, the generating
function for the observed degree sequence is given by
\begin{align*}
 g^{\obs}(z) 
& =  \sum_m a^{\obs}_{m+1}\,z^{m+1} \\
& =  z \sum_i a_i \,\sum_{m=0}^{i-1} z^{m} \Bigg[ \int_0^1 \,i t^{i-1} \,{i-1 \choose m} 
 \,p_\vis(t)^m \,(1-p_\vis(t))^{i-1-m}\,\dt \Bigg] \\
& =  z \sum_i a_i \Bigg[ \int_0^1 \,it^{i-1} \,\sum_{m=0}^{i-1} {i-1 \choose m} 
 (z p_\vis(t))^m \,(1-p_\vis(t))^{i-1-m}\,\dt\Bigg] \\
& =  z \sum_i a_i \int_0^1  \,it^{i-1} 
\left( 1-(1-z) p_\vis(t) \right)^{i-1}\,\dt \\
& =  z \int_0^1 \sum_i a_i  i \cdot 
\left[ t \left( 1-(1-z) p_\vis(t) \right) \right]^{i-1} \,\dt\\
& =  z \int_0^1 \,g'\!\left[ t \left(1-(1-z)p_{\vis}(t) \right) \right] \,\dt\\
& =  z \int_0^1  \,g'\!\left[ t - \frac{1-z}{g'(1)} \,g'\!\left( \frac{g'(t)}{g'(1)}\right) \right] \, \dt\enspace . 
\end{align*}
which completes the proof of Theorem~\ref{thm:genfunc}.

\medskip
Our definition of ``\reasonable'' degree sequences implies that the
graph is \whp\ connected, so that every copy is eventually added to
the queue.  For other degree sequences, Molloy and
Reed~\cite{molloy:reed:critical,molloy:reed:size} established that
\whp\ there is a unique giant component if $\sum_j a_j (j^2 - 2j) >
0$, and calculated its size within $o(n)$.  We omit the details, but
$g^\obs(z)$ is then given by an integral from $t_0$ to $1$, where
$t_0$ is the time at which the giant component has \whp\ been
completely exposed; this is the time at which $\cunto[t] = \cunex[t]$,
namely the largest root less than $1$ of the equation
\begin{align}
\label{eq:t0}
\sum_j j a_j t^j & =  t^2 \sum_j j a_j \enspace . 
\end{align}

\section{Examples}
\label{sec:examples}

\subsection{Regular graphs}
Random regular graphs present a particularly attractive application of the machinery
developed here, as the generating function for a $\delta$-regular
degree sequence is simply $g(z)=z^{\delta}$. From~\eqref{eq:gobs}, we derive
the generating function for the observed degree sequence:
\begin{align} \label{eq:gobsd}
g^{\obs}(z) 
& = z\delta \cdot \int_0^{1} t^{\delta-1}(1-(1-z)t^{\delta(\delta-2)}) ^{\delta-1} \dt \enspace .
\end{align}
This integral can be expressed in terms
of the hypergeometric function $\hg$~\cite{seaborn:hypergeometric-functions}.  
In general, for all $a>-1$ and $b>0$, we have
\begin{align*}
\int_{0}^{1} t^{a}(1-x t^{b})^{-c} \,\dt
& = \frac{1}{a+1} \,\hg\!\left(\frac{a+1}{b},~c;~\frac{a+b+1}{b};~x \right) \enspace .
\end{align*} 
where
\begin{align*} \hg(s,t;u;z) 
& =  \sum_{i=0}^\infty \frac{\Gamma(s+i)}{\Gamma(s)} 
\,\frac{\Gamma(t+i)}{\Gamma(t)}
\,\frac{\Gamma(u)}{\Gamma(u+i)}
\frac{z^i}{i!} \enspace,
\end{align*}
and $\Gamma(s+i)/\Gamma(s)$ is the rising factorial $(s)_{i} = s(s+1)(s+2)\dots(s+i-1)$, also known as the Pochhammer symbol.
In~\eqref{eq:gobsd}, $a = \delta-1$, $b = \delta(\delta-2)$, and $c=1-\delta$ (note $a > -1$ and $b > 0$ since $\delta > 2$) giving
\begin{align}
g^\obs(z) 
& =  z \cdot \hg\!\left( \frac{1}{\delta-2}, 1-\delta; 1+\frac{1}{\delta-2} ; 1-z \right) 
\label{eq:gobsint}
 \enspace .
\end{align}
Another useful identity is that for any negative integer $q$,
\begin{align*}
\label{eq:hgident}
 \hg(p,q;r;x) 
 = \frac{\Gamma(r) \,\Gamma(r-p-q)}{\Gamma(r-p) \,\Gamma(r-q)} 
\,\hg(p,q;p+q+1-r;1-x)
\enspace . 
\end{align*}
Here, $q = c = 1-\delta$, and $\delta$ is an integer greater than $2$.  
Thus,~\eqref{eq:gobsint} becomes
\begin{align*}
 g^{\obs}(z) 
& =  z \cdot \frac{\Gamma(1+\frac{1}{\delta-2}) \,\Gamma(\delta)}{\Gamma(\delta+\frac{1}{\delta-2})} 
\cdot \hg\!\left( \frac{1}{\delta-2},1-\delta ; 1-\delta ; z \right)  \\
& =  z \cdot \frac{\Gamma(\delta)~}{\Gamma(\delta+\frac{1}{\delta-2})~(\delta-2)} 
\sum_{m=0}^{\delta-1} \Gamma\!\left( m+\frac{1}{\delta-2} \right) \,\frac{z^m}{m!} \enspace ,
\end{align*}
where the summation now ranges over only the non-zero coefficients, \ie for all $m\geq\delta$, the rising factorial  term \mbox{$(1-\delta)_{m} = 0$} and the corresponding coefficients are zero. Thus, the expected observed degree sequence is given by
\Equation{a^{\obs}_{m+1}}{\frac{\Gamma(\delta)~\Gamma(m+\frac{1}{\delta-2})}{\Gamma(\delta+\frac{1}{\delta-2})~(\delta-2)~m!}\enspace.}
To explore the asymptotic behavior of $a^{\obs}_{m+1}$, note that 
\[ \Gamma(m) < \Gamma(m+\eps) < \Gamma(m) \,m^\eps \]
for all $m \ge 2$ and all $0 < \eps < 1$.  
Therefore, for $m \geq 2$, we can bound $a^{\obs}_{m+1}$ as follows:
\[ 
\frac{m^{-1}}{\delta^{1/(\delta-2)} \,(\delta-2)} 
\;\;<\;\; a^\obs_{m+1} 
\;\;<\;\; \frac{m^{-1+1/(\delta-2)}}{\delta-2} \enspace .
\]
For any fixed $\delta$, this gives a power-law degree sequence, and
in the limit of large $\delta$, one observes
$a^\obs_{m+1} \sim m^{-1}$.  Thus, even regular graphs appear to have a
power-law degree distribution (with exponent $\alpha \to 1$ in the limit $\delta \to \infty$) 
under traceroute sampling!

\subsection{Poisson degree distributions}
Clauset and Moore~\cite{clauset:moore:internet-mapping} used the
method of differential equations to show that a breadth-first tree in
the giant component of $G(n,p=\delta/n)$ has a power-law degree
distribution, $a_{m+1} \sim m^{-1}$ for $m\lesssim \delta$.  Here, we
recover this result as a special case of our analysis.  Recall that
\whp\ the degree distribution of $G(n,p=\delta/n)$ is within $o(1)$ of a
Poisson degree sequence with mean $\delta$.  The generating function is
then $g(z)=\e^{-\delta(1-z)}$, and the generating function for the observed
degree sequence is 
\begin{align}
g^{\obs}(z) 
& = z\delta \cdot \int_{t_{0}}^{1} \e^{-\delta(1-t)}\e^{-\delta(1-z) \e^{-\delta(1-\e^{-\delta(1-t)})}} \dt 
\nonumber \\
& = z \int_{0}^{1-t_{0}} \e^{-\delta(1-z)\e^{-\delta y}}\dy 
\enspace . 
\label{eq:gobsp}
\end{align}
In the second integral we transform variables by taking $y=1-\e^{-\delta(1-t)}$.  
Here, $t_0$ is the time at which we have exposed the giant component, 
\ie when $\cunex[t] = \cunto[t]$; since
$\cunex[t] = \delta t^2$ and $\cunex[t] = \delta t\e^{-\delta(1-t)}$, 
$t_0$ is the smallest positive root of $t = \e^{-\delta(1-t)}$.


This integral can be expressed in terms of the exponential 
integral function $\Ei(z)$~\cite{spanier:oldham:exponential-integral} and the
incomplete Gamma function $\Gamma(a,z)$, which are defined as
\begin{align*}
 \Ei(z) & =  - \int_{-z}^\infty \frac{\e^{-x}}{x} \,\dx \\
 \Gamma(a,z) & =  \int_z^\infty x^{a-1} \e^{-x} \,\dx \enspace .
\end{align*}
Then, with the integral
\[ \int_{p}^{q} e^{a \e^{b y}}\dy = \frac{1}{b} \left( \Ei\!\left[a \e^{q b}\right] - \Ei\!\left[a \e^{p b}\right] \right) \]
and the Taylor series
\begin{align*}
\Ei(-\delta(1-z)) 
& =  \Ei(-\delta) - \sum_{k=1}^\infty \frac{\Gamma(k,\delta)}{\Gamma(k)} \frac{z^k}{k}
\enspace , 
\end{align*}
taking $a = -\delta(1-z)$ and $b = -\delta$ as in~\eqref{eq:gobsp} gives\begin{align}
\label{eq:gobsp2}
g^{\obs}(z)
& \approx  \frac{z}{\delta}\, \left(\Ei\Big[-\delta(1-z)\Big] - \Ei\!\left[ -\delta\e^{-\delta(1-t_0)} (1-z) \right] \right) 
\nonumber \\
& =  \sum_{m=0}^{\infty} \frac{z^{m+1}}{\delta m!} \left( \Gamma(m,\delta \e^{-\delta(1-t_0)}) 
- \Gamma(m,\delta) \right) \enspace .
\end{align}
Thus, the coefficients of the observed degree sequence are
\begin{align}
a^{\obs}_{m+1} 
& =  \frac{1}{\delta m!} \int_{\delta \e^{-\delta(1-t_0)}}^\delta \e^{-x} x^{m-1} \,\dx
\enspace .
\label{eq:aobsp}
\end{align}
Now, $t_0$ approaches $e^{-\delta}$ in the 
limit of large $\delta$, and for $m \lesssim \delta$, the
integral of~\eqref{eq:aobsp} coincides almost exactly with the full
Gamma function $\Gamma(m)$ since it contains the peak of the
integrand.  Specifically, in~\cite{clauset:moore:internet-mapping} 
Clauset and Moore showed that 
if $m < \delta - \delta^\kappa$ for some $\kappa > 1/2$, then
\[ 
a^{\obs}_{m+1}
\;\;=\;\; (1-o(1)) \frac{\Gamma(m)}{\delta m!} 
\;\;\sim\;\; \frac{1}{\delta m} \enspace ,
\]
giving an observed degree sequence of power-law form $m^{-1}$ up to $m
\sim \delta$ and confirming the experimental result of Lakhina et
al.~\cite{lakhina:byers:crovella:xie}.



\section{Conclusions}
\label{sec:conc}

Having established rigorously that single-source traceroute sampling
is biased, thus formally verifying the empirical observations of
Lahkina et al.~\cite{lakhina:byers:crovella:xie}, and having calculated 
the precise nature of that bias for a broad class of random graphs, there 
are several natural questions we may now ask.

Petermann and De Los Rios~\cite{petermann:delosrios:exploration} and 
Clauset and Moore~\cite{clauset:moore:internet-mapping} both demonstrated experimentally 
that when the graph does have a power-law degree distribution, 
traceroute sampling can significantly underestimate the exponent $\alpha$.   
Although a characterization of this phenomenon is beyond the scope of this paper, 
it is a natural application of our machinery.

However, a more intriguing question is the following: can we invert 
Theorem~\ref{thm:genfunc}, and derive $g(z)$ from $g^\obs(z)$?  In other
words, can we undo the bias of traceroute sampling, and infer the
most likely underlying distribution given the observed distribution?
Unfortunately, it is not even clear whether the mapping from $g(z)$ to
$g^\obs(z)$ is invertible, and the complexity of our expression for
$g^{\obs}(z)$ makes such an inversion appear quite difficult. We
leave this question for future work.

Finally, although several studies claim that using additional sources
in mapping the Internet has only a small marginal
utility~\cite{barford:bestavros:byers:crovella,pansiot:grad:routers}, Clauset and
Moore~\cite{clauset:moore:internet-mapping} showed empirically that in
power-law random graphs, the number of sources required to compensate
for the bias in traceroute sampling grows linearly with the mean
degree of the network.  However, a rigorous analysis of multiple
sources seems quite difficult, since the events that a given edge
appears in BFS trees with different roots are highly correlated.
We leave the generalization of our results to traceroute sampling with
multiple sources for future work.

\subsubsection*{Acknowledgments}
A.C. and C.M. are supported by NSF grants PHY-0200909, ITR-0324845, CCR-0220070 and EIA-0218563, and 
D.K. was supported by an NSF Postdoctoral Fellowship.
We thank Jon Kleinberg, Andr\'{e} Allavena, Mireille Bousquet-M\'{e}lou and Tracy Conrad for helpful conversations.  
C.M. also thanks Rosemary Moore for providing a larger perspective.

\bibliographystyle{plain}

\end{document}